\documentclass{article}      
\input epsf

\def\b{\bigskip}
\def\bb{\bigskip\bigskip}

\def\r{\rightline}
\def\ce{\centerline}
\def\ve{\vfill\eject}

\def\r{\rightline}

\def\L{{\cal L}}

\def\harr#1#2{\smash{\mathop{\hbox to .25 in{\rightarrowfill}}
  \limits^{\scriptstyle#1}_{\scriptstyle#2}}}

\def\R{{\cal R}}

\def\today{\ifcase\month\or January\or February\or March\or April\or
May\or June\or July\or
August\or September\or October\or November\or  December\fi
\space\number\day, \number\year }

\r \today

\bb\bb\bb

\def\DD{\vec \bigtriangledown}

\def\w{\wedge}

\def\D{{\cal D}}
\def\A{{\cal A}}

\def\e{\rm e}

\def\p{\partial}

\def\sqr#1#2{{\vcenter{\vbox{\hrule height.#2pt
\hbox{\vrule width.#2pt height#2pt \kern#2pt
\vrule width.#2pt}
\hrule height.#2pt}}}}

  \def\square{\mathchoice\sqr34\sqr34\sqr{2.1}3\sqr{1.5}3}
\def\vac{|0\rangle}
  \def\1/2{{\scriptstyle{1\over 2}}}
  \def\a/2{{\scriptstyle{3\over 2}}}
  \def\5/2{{\scriptstyle{5\over 2}}}
  \def\7/2{{\scriptstyle{7\over 2}}}
  \def\3/4{{\scriptstyle{3\over 4}}}

\font\steptwo=cmb10 scaled\magstep2

\def\picture #1 by #2 (#3){
  \vbox to #2{
    \hrule width #1 height 0pt depth 0pt
    \vfill
    \special{picture #3} 
    }
  }

\def\scaledpicture #1 by #2 (#3 scaled #4){{
  \dimen0=#1 \dimen1=#2
  \divide\dimen0 by 1000 \multiply\dimen0 by #4
  \divide\dimen1 by 1000 \multiply\dimen1 by #4
  \picture \dimen0 by \dimen1 (#3 scaled #4)}
  }

 \usepackage{color}
\begin{document}

\def\sqr#1#2{{\vcenter{\vbox{\hrule height.#2pt
\hbox{\vrule width.#2pt height#2pt \kern#2pt
\vrule width.#2pt}
\hrule height.#2pt}}}}

  \def\square{\mathchoice\sqr34\sqr34\sqr{2.1}3\sqr{1.5}3}
\def\vac{|0\rangle}

\def\r{\rightarrow}
\def\M{{\cal M}} 
\def\D{{\cal D}}
\b
\def\DD{\vec\bigtriangledown}


\vskip-1cm
  {\ce {\steptwo  Relativistic Thermodynamics, a Lagrangian Field Theory}
  
  \ce{\steptwo
  for general flows including rotation} } 
\bb
  \ce{Christian Fr\o nsdal \footnote*{email: fronsdal@physics.ucla.edu}}
\b
  \ce{\it Physics Department, University of California, Los Angeles CA
  90095-1547 USA}
 \

 ABSTRACT ~~~Any theory that is based on an action principle has a much greater predictive power than one that does not have   such a formulation. The formulation of a dynamical theory of General Relativity, including matter, is here viewed as a problem of coupling Einstein's theory of pure gravity  to an independently chosen and well defined  field theory of matter. It is well known that this is   accomplished in a most natural way when both theories are formulated as  relativistic, Lagrangian field theories,  as is the case with Einstein-Maxwell theory. Special matter models of this type have ben available;  here a  more general thermodynamical model that allows for vortex flows is presented.

In a wider context, the problem of subjecting hydrodynamics and thermodynamics to an action principle is one that
has been pursued for at least 150 years. A solution to this problem has been known for some time,
 but only under the strong restriction to potential flows. 
 
 A variational principle for general  flows has  become available.
 It  represents a development of the Navier-Stokes-Fourier approach to fluid dynamics.
The principal innovation is the recognition that two kinds of flow velocity fields
are needed, one the gradient of a scalar field and the other the time derivative of a vector field
that is closely associated with vorticity. 
In the relativistic theory that is presented here  the latter is the Hodge dual of an exact 3-form, well known as the notoph field of Ogievetskij and Palubarinov, the $B$-field of Kalb and Ramond and the vorticity field of Lund and Regge. The total number of degrees of freedom  of a unary system, including the density and the two velocity fields  is 4, as expected.
  
 The present paper deals with the relativistic context, Special Relativity and General Relativity. The energy momentum tensor has a structure that is more general than that  
of Tolman, and different from  proposed generalizations; it appears to be well suited to represent rotational flows in General Relativity.  The current  of mass flow is conserved: the theory
incorporates the  hydrodynamical equation of continuity.  

\ve

\noindent{\bf I. Introduction}

The interaction between metric and matter is expressed by Einstein's field equation $G_{\mu\nu} = 8\pi GT_{\mu\nu}$. The model  for the right hand side that has been in use for 80 years is a phenomenological expression  suggested by Tolman. 
This paper begins with a review of the  phenomenological approach and  presents an alternative.
Thermodynamics, formulated as an action principle, is integrated with the Einstein-Hilbert action principle of General Relativity.  The source of Einstein's equation is the energy-momentum tensor of this relativistic field theory. The validity of the Bianchi constraint is thus assured in the  natural way.

  It was already known that, if it is  agreed to deal only with potential flows,  then  an integration of General Relativity with hydrodynamics is possible (Fronsdal 2007a).  
This theory was generalized to include Gibbs' principle of minimum energy  
(Fronsdal 2014a) and used for a general formulation of Thermodynamics as an action principle, but with the strong limitation to potential flows. A generalization that includes general flows has been proposed in a recent paper (Fronsdal 2014b). The kinetic part of the Lagrangian combines elements of `Eulerian' and `Lagrangian' hydrodynamics. It employs two vector fields,  the gradient of a scalar field, and the time derivative of a vector field. The natural and straightforward  joining of the two standard versions of hydrodynamics is a  development of the Navier-Stokes-Fourier 
approach to thermodynamics, including the standard treatment of viscosity and dissipation. The concepts are intimately 
related to work of Lund and Regge (1976). In addition to the potential velocity field associated with a type of flow around a vortex line they introduced a field $\vec x$, defined everywhere but incorporating the motion of the vortex line, as an additional dynamical field. We have proposed a role for this field in the general case of non potential flow.

\b
\ce{\bf The role of the equation of continuity}

\ce{\bf and the number of degrees of freedom}

Theories with strong predictive power are based on action principles; it is a curious fact that they do not include an equation of continuity. Is there a difficulty? Can we learn from it?

If the equations of motion are derived from an action principle, and if the equation of continuity is among them, then it must be related to the variation of a scalar field, the canonical conjugate of the mass density. The dynamical variables of the  simplest hydrodynamical systems are a density and the three components of the flow velocity, making a total of 4 variables. If the equations of motion are first order differential equations then there are four degrees of freedom altogether, or two canonical pairs.

Consequently, any action principle for a unary system in hydrodynamics must involve a velocity potential, and additional degrees of freedom to describe non-potential motion, and these addtional degrees of freedom must consist of  only one canonical pair.  The natural choice for an additional 
velocity variable is the time derivative $\dot{\vec X}$ of a vector field, since only in that case are the Euler-Lagrange equations of first order in $d/dt$.
This field appears to have  3 degrees of freedom, which does indeed suggest that there 
is an obstruction, but fortunately a solution is known
in the relativistic case and of course it has a non-relativistic limit. In the relativistic context of this paper the solution is the notoph field of  Ogievetskij and Palubarinov (1964). This field has only one propagating mode.
See Section VI.
\bb

\ce{\bf The nature of relativistic velocity}

The lifting of this theory to the relativistic context, which is the subject of this paper, has some dramatic elements. It has always been assumed that the velocity field of hydrodynamics is to be extended to a timelike vector field in the relativistic context. For fluids and solids this prescription was used by Minkowski (1908) to
deduce the constitutive relations of electrodynamics, and this procedure has been used in all the textbooks over a period of 100 years.   It is also a principal tenet on which Tolman based his formula for the energy momentum tensor; 
his formula for the energy momentum tensor has been generalized later but the flow has always been 
represented by a timelike four vector field.  Yet it is clear that the vector field of `Lagrangian' hydrodynamics, the time derivative of a vector field,  cannot be  the spatial part of a timelike 4-vector. 
The most economical way to imbed it into a relativistic theory is to
relate it to the Hodge dual of an exact 3-form. 

All of the physical insight that brought Tolman to suggest the formula
$$
T_{\mu\nu} = (\rho+p)U_\mu U_\nu - p\,g_{\mu\nu}
$$
for the energy momentum tensor to be used in gravitation is based on the motion of particles,
as is the further elaboration of the theory by Weinberg and others. But fluid flow is something else.
An example is the energy momentum tensor associated with potential flow of a gas, 
$$
T_{\mu\nu} = \rho\psi_{,\mu}\psi_{,\nu} - pg_{\mu\nu},
$$
this one derived from an action principle.  Much more dramatic is the effect of incorporating the
notoph field; it is not a time-like four vector and this kind of flow is apparently not related to the trajectory of particles, or to geodesics. This new velocity field is needed whenever non-potential motion - vorticity -  is involved. It implies
a structure of the energy momentum tensor that is very different from that postulated by Tolman. But the proposed change in outlook goes even deeper; it is a profound change
in the structure of `velocity space'. We are reminded that the first step in the formulation of a Lagrangian theory must be to review the choice of dynamical variables. 
\bb

\ce{\bf The case for action principles}

Modern astrophysics is not based on an action principle, although
Einstein's action has been invoked many times, after it was used by Eddington
in his book (Eddington 1926). 
 
We propose that the unification of thermodynamics with General Relativity that is needed for a theory of gravitation should  begin by formulating thermodynamics as an action principle. The
emphasis on variational principles for thermodynamics
is anything but new,  it was the inspiration and guiding light of  Gibbs' famous 300 page  paper (Gibbs 1878). 

It is significant that all unified theories studied so far are based on action principles, and this even applies to some proposals for combining Thermodynamics 
with General Relativity (Taub 1954, Bardeen 1970, Schutz 1970). 

The prototype of a unified field theory is the unification of the theories
of Maxwell, Einstein and Dirac that is defined by a combined Lagrangian.   But `electrodynamics' is a vast edifice that includes, 
besides the elementary interactions of photons and electrons,  
the field of electromagnetic interactions of fluids. The Maxwell-Einstein-Dirac system does not encompass electromagnetism in this widest sense; 
it is principally a relativistic theory of elementary particles. 

The potential benefits of formulating a fully fledged action principle 
for thermodynamics extend far beyond the present context; one important 
advantage is that each successful application places constraints on the
Lagrangian that  must be respected in subsequent applications.  In principle, one should aim at a ``Lagrangian for hydrogen" and a ``Lagrangian for  water" that would attempt to encompass diverse properties of those substances, in several phases.
It is less demanding, and thus less interesting, to account for individual phenomena, one at the time, using concepts and methods especially developed for each problem.

The imposition of  additional restrictions on any scientific program 
enhances the impact of the results of the investigation. Relaxation of constraints makes it too easy to account for observed phenomena and weakens the implications. 
\bb

\ce{\bf Thermodynamics}
 In the strictest sense thermodynamics is the study of equilibria, 
but of course we shall not be confined to that narrow
context. In the first place, to make meaningful contact with General
Relativity, we have to deal with a localized thermodynamics of fluids, including hydrodynamics. In the second place we shall have to  deal with systems with several components and several phases,  with chemical reactions, with flow 
and with a general class of equations of state. 

A fixed Lagrangian determines an adiabat and applies to adiabatic fluctuations. 
Among non-adiabatic changes the least difficult are quasi-static and 
develop on a very long time scale. Most treatments of continuous, non-adiabatic changes
assume that the time development proceeds through a sequence of adiabatic equilibria.
In the case of the simplest systems the adiabats may be indexed by a single parameter,
related to the entropy. A general expression for the Lagrangian, containing a free parameter,  applies to a family of systems interrelated by the change of entropy. 
This, together with the approximate additivity of Lagrangians of composite systems, is a promising starting point
for a study of some non adiabatic processes of heterogeneous systems.
The action principle provides a convenient framework within which to study entropy, the most difficult concept in thermodynamics, and dissipation.

\bb

\ce{\bf Electrodynamics}

In 1908 Minkoski  and, in the same year, Einstein and Laub, proposed constitutive relations for
the fields $\vec E,  \vec B,  \vec D$ and $\vec H$ in a fluid, or in a solid, subject to acceleration.
The idea is to begin with the case of a body at rest, for which the relations are known, and to
obtain relations for the case of a body moving with uniform velocity by using the transformations
of the special theory of relativity; this seemed natural since, in 1908 if not today, a direct confirmation of those transformation formulas was desired. Unfortunately, experiments to test the result were not carried out with uniform motion but with rotational motion; that is, accelerated motion (Wilson  1904, Wilson and Wilson 1913). This circumstance escaped comment and  the procedure  was solidly nailed down by 3 well known text books that appeared  during the 1960s. Incidentally, Feynman used the same procedure in his study of the laws of motion of superfluid Helium, 
with  a  comment on the danger of doing so (Feynman 1954 page 272). 
Only in 1995 did  Pellegrini and Swift (1995)  come forward with their severe criticism and definitive 
rejection of the theoretical analysis of 1908. That is certainly progress,  but the problem is that 20 later later there is still no analysis that is generally accepted.  A symptom of that is the surprising fact that 
something as  long established as the Lorentz transformation properties of the electromagnetic fields are now subjects of  debate.

Batchelor, in a celebrated paper  (Batchelor 1950) makes it very clear that something is needed. He says:
``Speculation about the explanation of these different physical phenomena would be less difficult and more profitable if there existed a background of general analysis concerning electromagnetic hydrodynamics, comparable with that which already exists for ordinary hydrodynamics."

A feature of  the analysis is, of course, that flow is represented by four-vectors. In view of recent developments that too may  have to be changed.

\b

\ce{\bf Superfluids}

If there  is one field where the number and the character of the degrees of freedom must
be in doubt it is that of superfluids.  It is a recurring theme, first introduced as Landau's phonons
and rotons, that several different flows are observed in rotating cylinders. In the early idiom,
vortices are phonons, dominated by potential flow except for the central vortex lines, and motions
generated by rotations of the outer cylinder are dominantly of the solid body type. In a more modern parlance there are two kinds of fluid, with two independent densities and two independent
velocities;  how many degrees of freedom must be invoked to describe and understand this interesting system?

In order to make this paper reasonably self contained, and to assure the reader that
the Lagrangian version of thermodynamics is well developed, it is necessary to devote  a part of it to an exposition of that theory.
 The subject is local thermodynamics and hydrodynamics, including systems with several components and phases, and with a general equation of state. As an example 
of astronomical applications we sketch a recent application to Dark Matter in the Milky Way. The reader may prefer to go directly to Section IV.
\b

\ce{\bf Summary}

Section II  reviews current phenomenological Relativistic Astrophysics and 
presents the original motivation for
the development of an action principle for Relativistic Thermodynamics.

Section III is a summary of recent progress towards such an action principle, limited to potential flow,  and Section IV lifts this  restricted formulation of thermodynamics to complete integration with General Relativity. An application of the theory to a problem of galactic dynamics is outlined.

 Section V describes a variational principle for thermodynamics that allows for general flows. The presentation is restricted to
simple, one component systems. A theory of general flows in mixtures has not yet
been developed. 

Finally, Section VI lifts this theory to a relativistic field theory, to offer a source for the dynamical metric field, consisting of matter with an arbitrary equation of state and with an unrestricted velocity field, suitable for a description of rotating objects in relativistic astrophysics. A most unexpected feature is that the velocity field is not a time-like four vector field. This signals a profound change in our understanding of hydrodynamics, thermodynamics and relativity.

\bb

\noindent{\bf II. The phenomenological approach}
\b

For the application of General Relativity to astrophysical and astronomical problems
Tolman (1934) proposed the following modification of pure gravitational dynamics,
$$
G_{\mu\nu} =  8\pi  G T_{\mu\nu}, ~~~T_{\mu\nu} = (\rho+p)U_\mu U_\nu - p\,g_{\mu\nu}.\eqno(2.1)
$$
Here $(G_{\mu\nu})$ is the Einstein tensor field derived from the Einstein-Hilbert action;  
 $\rho$ and $p$ are scalar fields and $U$ is a vector field; the only 
{\it a priori} restriction imposed on these fields is the normalization condition
$$
g^{\mu\nu}U_\mu U_\nu = 1.\eqno(2.2)
$$
The principal difficulty that we have with this suggestion is that it does not include
any dynamics for the matter sector, except for the normalization condition (2.2) that would be appropriate for the dynamics of non interacting particles.  For all the physics that went into constructing the expression, it is not the energy momentum tensor associated with any  known system.

The left  hand side of Eq. (2.1) satisfies the Bianchi identity; therefore consistency demands that the right hand side satisfy the `Bianchi constraint',
$$
{T^{\mu\nu}}_{;\nu} = 0.\eqno(2.3)
$$
Tensor fields that satisfy this constraint are the energy momentum tensors of dynamical, relativistic field theories, but Tolman's $T$, by itself, is a purely phenomenological expression, 
with no independent dynamics, and the constraint (2.3) is imposed in consequence of the coupling to the metric, not by our knowledge of a system that we want to interact with the metric.  Note that (2.3) is not a conservation law except in special circumstances.

An analogy helps to clarify this point.  Maxwell-Dirac electrodynamics has
$$
\p_\mu{F^{\mu\nu}} = J^\nu.
$$ 
The left hand  side is identically divergenceless and for this reason the current must satisfy the constraint ${J^\nu}_{,\nu} = 0$.  If no such current is available then one  resorts to phenomenolgy. In the case
of the problems with electromagnetism discussed in the preceding section we can invent a model for the matter system, but there can be no prediction of the outcome of an experiment. It is only
when we put strong limits on acceptible models that agreement with experiment has any meaning. A
 field theory of matter is qualified to interact with the vector potential only if it can offer a current that is conserved by virtue of the matter field equations. 
And a field theory of matter is qualified to interact with the metric only if it can present a tensor field that satisfies the Bianchi constraint by virtue of the  field equations that define it.

There are other  problems as well: the model offers no conserved mass density. Since the intention is to provide a relativistic generalization of 
hydrodynamics, where the equation of continuity is one of only two fundamental relations, it is a defect that is difficult to ignore. In fact, Weinberg (1972), and also Misner, Thorne and Wheeler (1972), have proposed to improve the prospects of the theory by introducing 
a second density $\sigma$ with the defining property that the current $\sigma U^\mu$ is conserved. It seems that this suggestion has not been taken up in astrophysical applications. 

The fact that Eq. (2.1) remains in use  in spite of these well known difficulties must be ascribed to a widespread belief that it may not be possible to improve on it. Of course the
expression for $T$ has been generalized.  
The literature is too vast to be quoted; but some relatively recent extensions
of Tolman's suggestion can be found in Israel and Stewart (1979), Andreassen (2011), Okabe  et al (2011), Schaefer and Teaney (2009). But the purely phenomenological nature of the approach has not  changed.  We feel that this approach may offer a useful description of a physical system but that it  has very little predictive power.  We shall see, in Sections III and IV, that a more dynamical approach is available in certain cases, and in Sections V and VI, that prospects for a more general solution of the problem are good.

Why turn thermodynamics, a phenomenological theory, into something more fundamental? 
Well, the problem before us is to combine thermodynamics with gravity; the alternative 
is to reduce General Relativity to phenomenology. Our main motivation is that any
results obtained will be more compelling when obtained under more restrictive conditions,
even in the case of restrictions that are mainly aesthetic. Here are two parallel
examples from theoretical physics.

1. The theory of electromagnetism in a continuous medium is particularly relevant to the discussion. Most textbooks (Jackson, Panofsky and Philips, Landau and
Lifshitz,...) take an approach that is parallel to that of Tolman. Instead of matter
dynamics, there is only a current that is subjected to the conservation law, required for Maxwell's equations to be solvable. Electrodynamics as a dynamical theory is thus compromised and reduced to phenomenology. This is by no means the only possible approach. 
Electrolysis, for example, can be formulated as a fundamental system that consists
of two charged fluids with opposite charge and a neutral component, with mutual interactions, subject to association/dissociation and minimally coupled to 
electromagnetism. And all this within an action principle. Whether such an approach is more practical  or not is not the issue: it is certainly more insightful and much more satisfying.

2. Of the two major studies of superfluid Helium one is fundamental, in the sense that it seeks an understanding of the underlying atomic theory
(Bogoliubov 1947). The other is more phenomenological (Landau and Lifshitz 1958, Feynman 1954 and others) but perhaps susceptible of  an action principle formulation.
Each approach throws some light on one or more aspects of the system and perhaps taken together they could provide a satisfactory, overall  picture. Instead, one or the other is invoked,
depending on the phenomenon under discussion. Within each context one strives to reach an elegant, dynamical and internally consistent formulation, but a common understanding is still far away.  At present it is usually assumed that Helium
below the $\lambda$ point is a mixture of two fluids, although Landau stressed an
alternative concept of two kinds of flow. In fact, `phonons' are associated with
potential flow and `rotons'  with solid-body motion.  The atomic picture does not recognize these concepte.

The preference for action principles was strong in Helmholtz, Maxwell and Einstein
and powerfully stated by Poincar\'e (1908):

``We cannot content ourselves with formulas simply juxtaposed which
agree only by a happy chance; it is necessary that these formulas come as it
were to interpenetrate one another. The mind will not be satisfied until it believes
itself to grasp the reason of this agreement, to the point of having the illusion that it could have foreseen this."

In as much as we are concerned with the relation between Thermodynamics and General Relativity it is natural to call attention to a recent paper by Verlinde (2009). This paper proposes an interesting connection between gravity and entropy,
but it does not develop a concrete approach to the problem of what is to be done with the matter contribution to Einstein's equations. The role of entropy in the
context of General Relativity is an important subject. The formulation of Thermodynamics as an action principle does much to advance our understanding of the entropy of heterogeneous systems (Fronsdal 2014c) and this advantage is retained in the passage to General Relativity.

 \ve

\noindent {\bf  III. Thermodynamics}
\b

 {\bf III.1. The energy functional of a simple system}

The dynamical variables of a unary, homogeneous system are the temperature, the pressure, the entropy and the volume.  

The  laws of classical, equilibrium  thermodynamics are expressed in terms of potentials $U, F, H$ or $G$.
Each is defined as a function of specific variables (the natural variables $T$ - temperature, $V$ - volume,  $P$ - pressure and $S$ - entropy), namely
$$
U(S, V), ~~F(T, V)~~H(S,P),~~G(P,T).
$$
According to Gibbs (1878), who quotes Massieu (1876), a thermodynamic system is completely defined by any
one of these functions; an explicit expression for a potential in terms of its 
natural variables is called a fundamental relation.  The four potentials are related by Legendre transformations and the fundamental relations for a given system are equivalent.

\b

The most fundamental concept introduced by Gibbs, enthusiastically promulgated by Maxwell,  is the representation of the states 
of a system in terms of surfaces in a Euclidean space with coordinates $ V,T,P$, $S$. Gibbs based the entire theory on a principle of minimal energy or on an equivalent principle of maximum entropy.  An energy functional is subject to  independent variations of 2 of the  variables,
the other 2 held fixed. The Gibbsean surfaces are 2-dimensional surfaces in four dimensions,
determined by the variational equations.
Physical configurations are points on this surface.

The action is thus a function of the four independent variables. To define the system one specifies the potential and
obtains 2 independent relations among the variables that define the Gibbsean two dimensional surface of physical states. 

The relations obtained by variation of the generic action must have universal validity and the following definition of  the energy functional satisfies that criterion,
$$
E = F(V,T) + ST +  V P.
$$

According to Gibbs,  in the description of adiabatic processes the energy is to be varied with entropy and pressure held fixed. Independent variation of $ V$ and $T$ give the standard relations
$$
{\p F\over \p T} + S = 0,~~ {\p F\over \p  V}+P = 0.
$$
An adiabatic system is thus defined by the free energy  function and the values chosen for $P$ and $S$.  To 
encode the properties of the system in the expression for the free energy $F$, rather than $U$ or $G$, is  often the preferred choice, see for example Rowlinson (1959).

Pavlov and Sergeev (2008), among others,  have discussed the interpretation of Gibbs' thermodynamics in terms of differential geometry.
In the fully developed context of `local' thermodynamics the dynamical variables
are fields over an infinite dimensional configuration space. The Gaussian manifolds are
infinite dimensional and the development of a canonical structure is interesting.
See Morrison (1984) and Marsden and Weinstein  (1982). This last paper is a good
source for references to the origins of the subject.  A glimpse of the theory is offered below, see Eq.(3.2) and (3.13).

\def\H{{\cal H}}

\bb

{\bf  III.2. Localization}

The local extrapolation of thermodynamics seeks to promote these relations to field equations that describe spatial but,  in the first instance, time independent configurations. Until further notice we consider systems with one component and  a single phase.
The function $F$ and the variable $S$ are interpreted as specific densities, the variable $ V$ as a specific volume. The mass density is $\rho = 1/ V$ and densities $f,s$ are defined by  
$$
f(\rho ,T) = \rho F(\rho,T),~~ s = \rho S.
$$
(The representation of the entropy density $s$ as a product $\rho S$ is meant to imply that the specific entropy density $S$, rather than $s$,  is to be kept fixed when the energy functional is varied.)   The energy density, now including the kinetic energy,  is
$$
h = \rho\vec v\,^2/2 + f(\rho,T) + sT,  \eqno(3.1)
$$
with $\rho, S$ and $T$ treated as independent variables.  The vector field $\vec v$ is interpreted as a flow velocity.     
 \b

We are going to extend the action principle to incorporate the hydrodynamic equation of motion for the velocity field, as well as the equation of continuity. We begin by introducing the Poisson bracket; for two functionals $f$ and $g$ over an algebra of dynamical field variables, differentiable and with suitable behavior at infinity,  
 $$
 \{f,g\} := \int d^3x \Big({\p f\over \p v_i}\p_i{\p g\over \p \rho} - {\p g\over \p v_i}\p_i{\p f\over \p \rho}\Big).\eqno(3.2)
 $$
 {\bf Remark.}  The fields $\rho$ and $\vec v$ are not natural variables. The formula takes a more conventional form below, when we specialize to potential flows by setting $\vec v = -\DD\Phi$, in Eq. (3.13) .  But even in this form it satisfies the Jacobi identity, since the mapping $f\mapsto \{f,g\}$ is a derivation.
 
\b 
 
\noindent The two fundamental equations of hydrodynamics take the form
 $$
 \dot\rho = \{H,\rho\} =- \DD \cdot (\rho\vec v)\eqno(3.3)
 $$
 and 
 $$
 \dot {\vec v} = \{H,\vec v\} = -\DD ~{\p h\over \p \rho}.\eqno(3.4)
 $$

\noindent The first equation is the equation of continuity and the second equation will be related to the Bernoulli equation.   The energy functional as defined by Eq.(3.1) is thus a fully fledged Hamiltonian.
 
\bb

{\bf III.3. The Lagrangian}
 
Going a step further, we pass to a Lagrangian formulation.
In the present context (but see Section V),  this can be done only if the velocity field is  irrotational, locally expressible as
 $$
 \vec v = -{\rm grad \,\Phi}.
 $$  
The correct form of the kinetic part of the action for hydrodynamics  was presented by Fetter and Walecka (1980).  Generalizing to thermodynamics, we  define an action and a Lagrangian density by
$$
{\cal A}=\int dt \int_\Sigma\,d^3x(\L-P)
$$
and
$$
 \L = \rho\,\dot\Phi -h,~~h = \rho(\vec v\,^2/2 +\phi) + f(\rho,T) + sT.\eqno(3.5)
$$
This makes $\Phi$ canonically conjugate to the density $\rho$.
The equations of hydrodynamics are now variational equations.
Variation of $\Phi$ and $\rho$ gives the Hamiltonian equations of motion (3.3-4) in the form
$$
\dot\rho = -{\p h\over \p \Phi}, ~~\dot\Phi = {\p h\over \p \rho}. 
$$
We have included, in (3.5), the gravitational potential $\phi$ of a fixed gravitational field.
 The scope of the theory is revealed by listing the Euler Lagrange equations:

$\bullet$ Variation of the field $T$ gives the adiabatic relation for densities,
$$
{\p f\over \p T} + s = 0.\eqno(3.6)
$$
For an ideal gas, in the case that the specific entropy density $S=s/\rho$ is uniform, this is the usual polytropic condition. In the case of multi component systems it has a significance that by far exceeds what is usually recognized.

$\bullet$ Variation of the velocity potential gives the equation of continuity, Eq.(3.3).
 
$\bullet$ Local variation of $\rho$  (variations that vanish at the boundary of the domain $\Sigma $), with $S$ and $P$ held fixed, gives 
$$
\dot\Phi -{\vec v}^2/2 -\phi - \mu = 0,\eqno(3.7).
$$
where $\mu$ is  the chemical potential, defined by either of two equivalent 
equations,
$$
\mu = {\p f\over \p \rho}\bigg|_T,\eqno(3.8)
$$
or
$$
\mu := {\p (f+sT)\over \p\rho}\Big|_S.\eqno(3.9)
$$
In the second form appears the adiabatic derivative, where $T$ is to be eliminated in
favor of $\rho$ and $S$ using (3.6).
 
A simple equation relates the gradient of the chemical potential to the gradient of the pressure,  defined by
$$
p= \rho{\p f\over \p \rho}-f,\eqno(3.10)
$$
namely
$$
\DD p = \rho\DD q,
$$
but the derivation is interesting. It follows from the identity
$$
\DD p -\rho\DD q =  (s-\rho{\p s\over \p\rho})\DD T 
-\rho T \DD {\p s\over \p \rho}.\eqno(3.11)
$$
The first term on the right is zero when the entropy density $s$ is linear in $\rho$;
the second term vanishes when the specific entropy $S=s/\rho$ is uniform. Both 
conditions are usually assumed, some times tacitly, to hold. In this case,
and only in this case, the Euler-Lagrange equation (3.7) is equivalent to the
Bernoulli equation,
$$
\rho{D\vec v\over D t} + \rho\DD\phi = -\DD p. \eqno(3.12)
 $$

 $\bullet$ A complementary variation of  the density, with the mass fixed but the volume not,
 gives the result that the thermodynamic pressure must coincide with the pressure $P$ at the boundary. In the interior the thermodynamic pressure is balanced by external forces.

\b

This thermodynamic action principle, based on the hydrodynamic Lagrangian of Fetter and Walecka, and on Gibbs' principle of minimum energy,  is adaptable to  incorporation  into the framework of General Relativity, as shall be shown in Section IV of this paper.  That is not the only advantage of this formulation of thermodynamics. The following sections give a very short account of some other applications and generalizations.

\b

 The Euler-Lagrange equations 
apply to adiabatic dynamics only. Relaxation is non adiabatic and irreversible.
What gives adiabatic dynamics its importance is the fact (the assumption)
that relaxation takes place on a much longer time scale. The variational equations determine the temperature as a function of the other variables, but there is no equation
for the time derivative $\dot T$. Over short time intervals the temperature follows the
variations of the density, over long time intervals it is determined by the heat equation.
To go further the paradigm that is outlined at the end of Section V may be followed.
 
The independent dynamical variables are now the scalar fields $\Phi, \rho$ and $T$, and the Poisson bracket can be expressed as
$$
\{f,g\} := \int d^3x\big({\p f\over \p \Phi}{\p g\over \p \rho}-{\p g\over \p \Phi}{\p f\over \p \rho}\big).\eqno(3.13)
$$
It is singular as seen by the absence of the variable $T$; the adiabatic condition is thus a constraint.

\b

Fig.1. is an attempt to present the Gibbsean view.  The parabolas represent adiabatic families of configurations, each characterized by a fixed value of the entropy. The horizontal line
is the manifold of adiabatic equilibria; a process of slow dissipation can best be
understood when it takes place along this line.

 \vskip2in  

\vskip-3cm
 
\epsfxsize.6\hsize
\centerline{\epsfbox{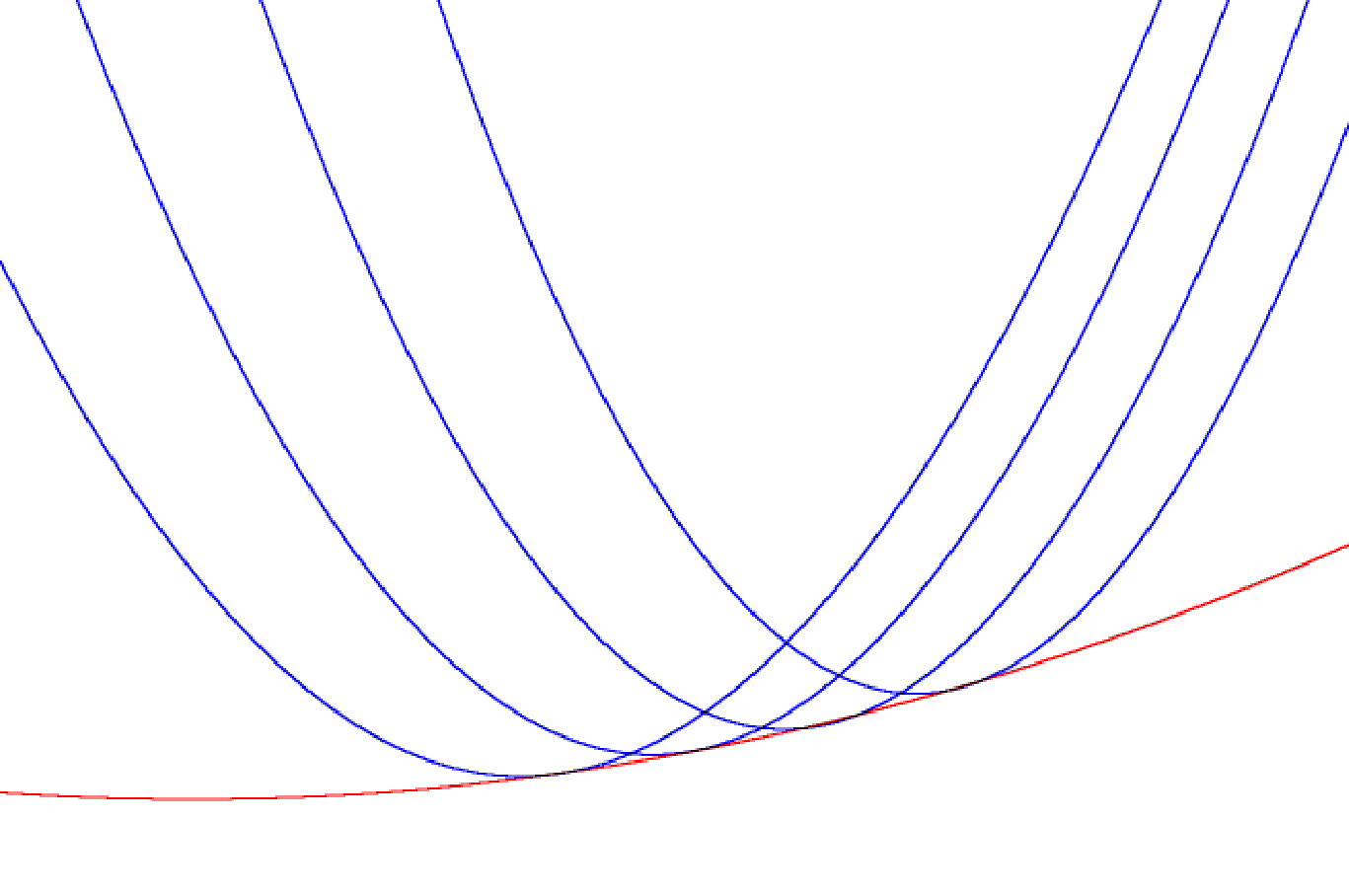}}
\vskip-.5cm

\parindent=1pc
\bb
Fig. 1. The  configurations of an adiabatic system (blue) form a  surface in the  Gibbsean   space of
thermodynamic  configurations. The collection of equilibria (red) form the space of classical
thermodynamics.  
\bb

\noindent{\bf Remark.} We shall see that the theory of Fetter and Walecka is the non relativistic limit of a Lorentz invariant Lagrangian field theory. It follows that it is invariant under the group of Galilei transformations. The representation that 
appears in this connection is known as a ``Galileon"; it has recently been the
subject of much mathematical interest. See for example  de Rham and Tolley (2010), Fairlie (2011), Curtright and Fairlie (2012). 
\bb

\noindent {\bf III.4. Generalizations}

Here we describe a fairly  general situation that can be encompassed by the theory in the present stage of development.

Consider a system described by two independently conserved densities,  two
independent, irrotational vector fields, temperature, entropy and external pressure. We suppose that we have an expression for the free energy density $f(\rho_1,\rho_2,T)$ and propose the following Lagrangian density
$$
\L =  {\sum_j } \rho_j (\dot \Phi_j - \vec v_j\,^2/2-\phi ) - f(\rho_1,\rho_2,T) -
T\sum_j^n\rho_jS_j +{\hat a\over 3}T^4.\eqno(3.10)
$$
We have included the Newtonian potential $\phi$, in recognition of the fact that the  
$\rho$' s  are interpreted as mass densities, and the Stefan-Boltzmann pressure, reserving comments  for Section IV.
 A velocity potential $\Phi_j$ is canonically conjugate to the conserved density $\rho_j, j = 1,2.$  
The central problem is the determination of the total free energy density.
The most primitive guess and a good approximation in some cases    
 is obtained by summation over the components,
$$
f(\rho_1,\rho_2,...;T) = \sum_if_i(\rho_i,T).\eqno(3.11)
$$
Corrections, in the form of interaction terms, are necessary and will be discussed.

If a chemical reaction is taking place only the total density is conserved and the 
first sum in (3.10) is reduced to one term. (The case when more than 2 components are present,
and one or more chemical reactions are taking place will not be discussed here
(Fronsdal 2014c)). The penultimate term in (3.10) represents a modeling of the entropy of mixtures, a new direction in the theory of mixtures.
The specific densities $S_1, S_2$ are usually taken to be uniform;
that is, spatially constant. The appearance of an arbitrary constant  in any adiabatic Lagrangian is expected, for it must be possible to add heat and this must be reflected in the equations of motion. But the appearance of  $n$ arbitrary constants (one for each component) requires that we specify the path that the system will follow in this $n$-dimensional ``entropy space".  In certain circumstances the  Gibbs-Dalton hypothesis may serve to fix the ratios, but a deeper analysis is
required to account for properties of real gases. For a system with chemical reactions, as in the case of the molecular dissociation of hydrogen,  the addition of heat to the compound system leads along a path on which $T(S_1-S_2)$ is fixed.   In the case of a saturated van der Waals gas, the path is defined by the Maxwell rule of equal areas, derived from the subsidiary principle of maximal entropy, with the result that $T(S_2-S_1) = \epsilon$, the evaporation energy.  In this case too,  non adiabatic (but reversible) changes are taking place in which the entropy is
changing and both cases provide some insight into the use  and the concept of entropy.
The representation of the total entropy density as a linear function of the mass densities is an attempt to model the entropy of mixtures; it is a novel and promising
idea that is natural in the Lagrangian context.

\bb

\ce  {\bf Example, hydrogen atmospheres}

Hydrogen is  present in most stars, in  the molecular form at low temperatures and the atomic form at higher temperatures.
Here we consider the phenomenon of molecular dissociation,   below
the ionization temperature. The Hamiltonian density, in a first approximation, is,
with $\rho := \rho_1 +\rho_2$,
$$
h = \rho( \lambda + \vec v\,^2/2 + \phi) +  T \sum_i\R_i\rho_i\ln{\rho\over T^{n_i}} +\sum \rho_iS_i T.
$$ 
The index is 1 for the molecular component and 2 for the atomic component.
  Variation of  the densities,
with $\rho =  \rho_1 + \rho_2$ fixed, leads to the following equation for the concentration $ r = \rho_2/\rho$ at fixed density
$$
{r^2\over 1-r} = {1\over e \rho}T^{.5}\e^{-\epsilon/\R T}.
$$
This is Saha's equation (Saha 1921). It was derived under the assumption that $S_1 - S_2 = 
\epsilon/T$, where $\epsilon$ is the molecular binding energy. 
\bb

\ce {\bf Phase transitions, speed of sound and mixtures}

For a pure substance some simple phase transitions and the associated dynamics are described by using the free energy of a van der Waals gas. As the temperature
is lowered (non adiabatically)  to the domain where the density is no longer determined by temperature and pressure the system behaves in some respects like a mixture, but surface tension and gravity leads to separation of the phases.   The theory describes the separated equilibrium configurations and adiabatic dynamics
but not the process whereby the system gives up or gains heat from its surroundings,
nor the process of separation of the phases. In the saturated, separated  phase
the total Lagrangian is simply the sum of the Lagrangians of the components, and
$$
f(\rho_1,\rho_2,...,T) = \sum_i f_i(\rho_i,T).\eqno(3.12)
$$
Variation of the densities gives the pressure and locates the critical point but it does not yield Maxwell's rule.   Maxwell's rule can be derived from Gibbs complementary variational principle of maximal entropy.

The question of the validity of Eq.(3.12) is vital in other contexts, such as the propagation of sound and the determination of critical parameters in gaseous mixtures. The latter problem is often formulated in terms of the free energy, while sound speeds are derived from very different considerations  (de la Mora and Puri 1985.) This brings us to what is without a doubt the most important consequence of adopting the action principle for all applications  of thermodynamics: Every successful application gives information about the Lagrangian. For example, the study of sound propagation in mixtures 
at normal temperatures suggests corrections to the formula (3.12) by interaction terms of the very specific form (Fronsdal 2014c)
$$
f_{\rm int} = \sum \alpha_{ij} (\rho_i\rho_j)^k,~~~k=.5.
$$
The most complete account of critical parameters of mixtures have been obtained by a
modification of the van der Waals formula for the pressure, but  preliminary
results based on the following formula for the free energy are also encouraging, 
$$
f(\rho_1,\rho_2,...;T) = \sum_if_i(\rho_i,T)  +\alpha_{ij} (\rho_i\rho_j). \eqno(3.13)
$$
Here $f_1, f_2, . . .$ are the van der Waals free energies of pure substances; the only adjustable parameters are the interaction parameters.
The particular form of the interacting term is suggested by  observation
of the excess free energy. When more data on sound propagation at low temperatures become available it will be imperative to reconcile observations
of critical phenomena with sound speeds in mixtures, using the same expression for the free energy to account for two very different phenomena.

The ultimate goal of thermodynamic phenomenology should be to determine the Lagrangian for each substance, including mixtures of all kinds and including all phases,
to account for a large family of adiabatic properties.

\bb

\noindent{\bf IV. General Relativity.  Potential flows.}

Following Taub (1954), Schutz  (1970) and Bardeen (1970), we have proposed a partial unification of thermodynamics and General Relativity.  But unlike previous suggestions this one is  defined by a specific Lagrangian density,
$$
\L_{tot} = {1\over 8\pi G}(R + \Lambda) + \L_{\rm mattter},
$$
where the second term is a relativistic extension of the Lagrangian density  in Eq.(3.5):
$$
\L_{\rm matter} = {1\over 2}{\sum_j}'\rho_j\big(g^{\mu\nu}\psi _{j,\mu}\psi_{j,\nu}-c^2\big) - f (\rho_1,\rho_2,...,\rho_n;T)- T\sum_1^n \rho_iS_i + {\hat a\over 3}T^4.
\eqno(4.1)
$$
(The simplest example of this theory was proposed by Fronsdal (2007a, 2007b).)

In as much as the on shell value of this density is equal to the negative pressure, this is the Lagrangian  proposed by Taub (1954).
The radiation term ($\hat a$ is the Stefan-Boltzmann constant)  may perhaps be included in the free energy, but we prefer to add it explicitly.  Eq.(4.1)  is expected to be valid for mixtures with no chemical interactions and well away from phase transitions.   The constant $\Lambda$ is the cosmological constant and in the case of an astrophysical system  that extends to infinity it is interpreted as the pressure at the boundary. 

Eq.(4.1) is a simple extension of the non relativistic Lagrangian (3.5) or (3.10) to which it reduces when in the kinetic term one expands
$$
\psi_i = c^2 t + \Phi_i
$$
and takes the limit as $c$ tends to infinity.
Of the dynamical metric there remains, to this order, only the Newtonian potential $\phi$. Up to terms of order $1/c^2$, 
$$
{1\over 2}(g^{\mu\nu}\psi_{i,\mu}\psi_{i,\nu} - c^2) 
= \dot\Phi_i -{1\over 2}(\DD\,\Phi_i)^2 -\phi.
$$

 The action is relativistically invariant (invariant under diffeomorphisms) and it follows that the Bianchi constraint is satisfied by virtue of the equations of motion.
   \b
 The matter energy momentum tensor is
 $$
 T_{\mu\nu} =  \sum_i \psi_{i,\mu}\psi_{i,\nu} -\L g_{\mu\nu}.
 $$
Hence $\L$, on shell, is the pressure,  just as it is in the non relativistic theory.  The expression is similar to that of Tolman (1934), with some differences:
\b

1. Any number of densities and (gradient) vector fields can be accommodated, with any equation of state.

2. The charges that are conserved in the non relativistic theory remain conserved in General Relativity,
$$
\p_\mu \, J_j^\mu = 0, ~~J_j^\mu = \sqrt{-g}g^{\mu\nu}\psi_{j,\nu}.
$$

3. The Lagrangian is determined by the non relativistic free energy and  interpretated  as in  the non relativistic theory. Simple chemical reactions and changes of phase can be described precisely as in non relativistic thermodynamics.

4. All the equations  of ordinary thermodynamics and hydrodynamics are recovered in the non relativistic limit. 

5. The simplest examples  (ideal gas spheres, no radiation) lead to equations of motion that are almost identical to those used in simple stellar models.

6. The  black body radiation pressure  and energy is included by simply 
adding the term $\hat a T^4/3$ to the Lagrangian density. This is appropriate in a region where radiation is  in equilibrium with matter, as in the photosphere of the Sun. It effectively  raises or lowers  the  adiabatic index towards the 
asymptotic limit $n = 3$ as the temperature goes to infinity. It contrasts with the approximation introduced by Eddington (1926) who postulated that the ratio $p_{\rm gas}/p_{\rm radiation}$ is uniform.
 
7.  The temperature is an independent field, fixed by the equation of motion, Eq.(3.6). This is important  in the case that radiation is taken into account for the
 polytropic relation is affected.  It leads to a correct inclusion of radiation pressure and energy.    
  
 8. The  pressure is the on-shell value of the Lagrangian density, so that the term $-pg_{\mu\nu}$ in the energy momentum tensor is the same in both theories, but the dynamics is in the expression for the Lagrangian and it is lost when the Lagrangian  is replaced by its value on shell. As to the other term,  it can be compared to Tolman's expression only in the case of a simple system. There is no conserved energy density  in General Relativity, but in the non relativistic limit
 $$
 \rho\dot\psi \dot\psi = \dot\psi{\p \L\over \p \dot\psi}  =h + \L = h+ p,
 $$
where $h$ is the Hamiltonian density. 

9. The entropy enters in precisely the same way as in the classical theory. Any 
advance in the conceptual or operational meaning of entropy is transferred to the relativistic theory, without additional complications.

10. Since the mass flow current is conserved, minimal coupling to the electromagnetic potential is possible and the theory then becomes a model of  a  charged relativistic plasma. This also allows to make contact with the Reissner Nordstr\"om solution of Einstein's equation 
(Fronsdal 2007b).
\b

The non relativistic theory of binary mixtures is easily lifted to the relativistic context, but a relativistic theory of more complicated mixtures has not yet been developed. 

The important problem of rotational and vortex motion in General Relativity has had to await the development of a variational approach that allows for irrotational flows, 
see the Section V.
\b

\ce  {\bf Dark matter, an example}

An interesting equation of state for stellar interiors, first proposed by Stoner (1930) and Chandrasekhar (1935), is based  on a degenerate Fermi gas.  A variant was used in a celebrated  paper by Oppenheimer and Volkov (1939) and an account of it may be found in Landau and Lifshitz (1958). Another variant of this equation of state was used recently to model the dark matter in the Milky Way  (Fronsdal and Wilcox 2011).  

Let $f$ be the internal energy density, and let us assume  that the entropy is zero;
 $f$ is effectively a function of the density. Variation of the action with respect to the density gives, in the case of a stationary, spherically symmetric metric,
$$
{1
\over 2}(c^2g^{00}-1) =: \phi = {d f\over  d\rho}.
$$
Taking the gradient leads to the usual hydrostatic conditions relating $\rho\,{\rm grad}\,\phi$ to the gradient of the pressure
$$
p = \rho{\p f \over \p \rho}-f.
$$
The gravitational potential is identified with the chemical potential. and
$$
\rho = {dp\over d\phi}.
$$
 Examples studied by Fronsdal and Wilcox (2011)  have (omitting constant factors)
 $
 \rho(\phi) =  \sinh^4\phi.
 $
 The non relativistic approximation leads to the Laplace equation, $\Delta\phi = \rho$,
 $$
 r^{-2}\p_rr^2\p_r \phi = \sinh^4\phi,
 $$ a ``sinh-Emden equation" that   is solved exactly by
 $$
 \phi(r) = \ln(1+ b/r).
 $$
It can be used to account for the observed rotation curve of visible stars in the Milky Way. It predicts nearly constant orbital velocities up to the distance $r = b$. The solution is regular all the way to the center and reveals a small core that resembles a black hole. A more elaborate model involving a mixture of two components gives 
an equation of state that can account for the high gravitational fields observed at
about $10^{16}$  cm from the center.  The presence of a non vanishing mass density appears to prevent the formation of black holes but to permit very massive 
accumulations of dark matter. (The equations of state used in this example  are not applicable to  standard, visible matter.)

\bb

\ce{\bf Two-phase stars}

The formalism makes it easy to deal with a gaseous star governed by the van der Waal's equation.
Boundary conditions at the gas/liquid surface are determined as usual by Maxwell's rule and the
solutions for the density and for the metric field are regular from the center to infinity. 
The density and the specific entropy are discontinuous at the surface.
A Black Hole may be formed in the limit when the atmosphere has condensed completely.

\bb

{\noindent V. {\bf Action principle for unrestricted velocity fields}}

  Action principles have been widely used in many branches of physics.  The fundamental work of Gibbs on heterogeneous systems is based on principles of minimum energy (and maximum entropy), but the actual construction of a Lagrangian for all but the simplest systems has not been a popular direction of research.  The original suggestion of Fetter and Walecka (1980), within the context of Eulerian hydrodynamics is successful,
but fundamentally limited to potential flows. For the same reason the recent development of this theory to cover a large part of thermodynamics has received scant attention.

A generalization, an action principle for thermodynamics that allows for general flows,
has been found by a detailed examination of the way that the 
Navier-Stokes equation accounts for certain kinds of laminar, rotational flows. A
full account of this work will  be published elsewhere (Fronsdal 2014b). Here, to control the length of this paper, we shall only describe the result. This course is also indicated
because the offered solution is quite natural and best evaluated on its own merits.

Hydrodynamics is familiar in two different classical formulations. The Eulerian version
has already been discussed in this paper.  It becomes a Lagrangian field theory
(an action principle) when the flow velocity is restricted to be a gradient,
$\vec v = -\DD \Phi$, and a direct consequence of this is that variation of the 
action with respect to the scalar field $\Phi$ gives the equation of continuity.
This field is conjugate to the density and is needed in any formulation of hydrodynamics.

In the alternative, `Lagrangian' formulation of hydrodynamics the velocity field
is represented as a time derivative, $\vec v = d{\vec X}/dt$.  (The more common notation $\vec v = d{\vec x}/dt$  suggests  the interpretation of $\vec x$ as
the position coordinate of a fluid element, but $\vec X$ is above all a field.) The fundamental variables are not $\rho$ and $\dot{\vec X}$ but $\rho$ and $\vec X$.
The Lagrangian is constructed from these variables  and variation of $\vec X$ gives
a vector equation. Here there is no need for $\dot{\vec X}$ to be irrotational, but the
price to be paid is that there is no equation of continuity.
\vskip.5cm

The new proposal is to combine both theories,
 $$
\L = \rho\Big(\dot \Phi + \dot{\vec X}^2/2 + \kappa \dot{\vec X}\cdot\DD\Phi-(\DD \Phi)^2/2\Big)-f-sT.\eqno(5.1)
$$
The conserved flow is 
$$
\vec J = \rho \vec v,~~ \vec v := \kappa\dot{\vec X} - \DD \Phi.\eqno(5.2)
$$
The principal advantage of this model is that it offers a first integral, in the form of the Hamilton   $H=\int d^3x \,h $, with the density
 $$
h = \rho\Big(\dot{\vec X}^2/2 +(\DD \Phi)^2/2\Big)+f+sT.\eqno(5.3)
$$
and an associated expression for energy flow. 

The Euler-Lagrange equations are: from variation of $T$ the adiabatic condition, from variation of $\Phi$ the equation of continuity,
$$
\dot\rho +\DD\cdot(\rho\vec v)=0,~~~\vec v := \kappa\dot{\vec X}-\DD\Phi,\eqno(5.4)
$$  
from variation of $\vec X$,
$$
{d\over dt}\vec m = 0,~~~ \vec m := \rho(\dot{\vec X}+ \kappa\DD\Phi),\eqno(5.5)
$$
and from variation of $\rho$,
 $$
 \dot \Phi + \dot{\vec X}^2/2 +\kappa \dot{\vec X}\cdot\DD\Phi-(\DD \Phi)^2/2-\mu =0.\eqno(5.6)
$$
They have the familiar structure of a set of conservation laws; (5.4) expresses mass conservation, (5.5) can be interpreted as momentum conservation and (5.6) with the rest
implies the conservation of energy. 

In this context the standard approach to dissipation is to add sources to one or more of the conservation equations. The 
system becomes dissipative when (5.5) is replaced by
$$
{d\over dt}\vec m = \hat \mu\Delta  \vec  v,\eqno(5.7)
$$
where $\hat \mu$ is the dynamical viscosity.  
This, as we shall see, brings us very close to the physics associated with the Navier-Stokes equation. The important difference is that here we have a canonical expression for the energy density and a unique  expression for the rate of energy dissipated by viscocity, namely
$$
{d\over dt} h = \mu\dot{\vec X}\cdot \Delta\vec v.
$$ 

Conserved energy is not a feature of the Navier-Stokes approach, even in the limit of no viscosity, and we are not comfortable with the practice of postulating an  {\it ad hoc} expression for ``energy". 

\b
\ve

\ce{\bf Application to laminar, stationary rotational flow}

We shall call it Couette flow, although this term invokes the various phenomena that
occur at higher rotational speeds.
Consider a situation that is effectively 2-dimensional because of translational symmetry, as when  nothing depends on a vertical coordinate $z$ and the 
direction of flow is confined to the horizontal $x,y$ plane. In cylindrical Couette flow the fluid is confined between two concentric cylinders that can be rotated around a common axis as in Fig.2.
With both cylinders at rest we postulate a state with all variables 
time independent  and uniform. The effect of gravity will be neglected. The cylinders are long enough  so that end effects can be neglected as well. The boundary conditions are no-slip.

\epsfxsize.5\hsize
\centerline{\epsfbox{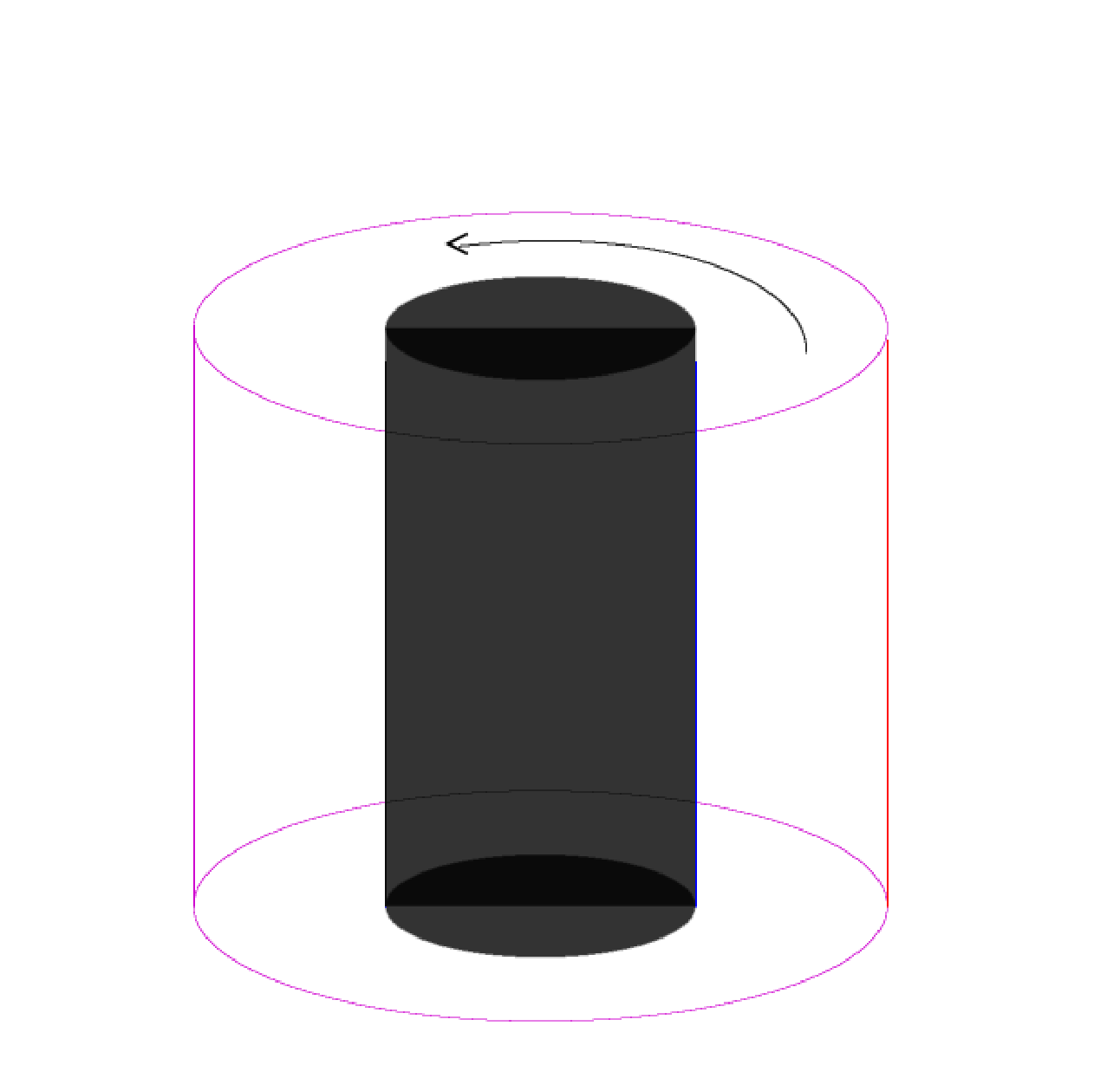}}
\vskip0cm

Fig.2. Cylindrical Couette flow.
\b
  
The only stationary circular flow (all variables, including both velocities but not the potentials $\Phi$ and $\vec X$,  are time independent), that is locally irrotational is
$$
\vec v =  - \DD  \Phi,~~~ \Phi :=  a\arctan {y\over x}
$$
or in the Cartesian basis (and $r := \sqrt{x^2+y^2}$) ,
$$
\vec v = {a\over r^2}(-y,x,0),~~~a={\rm constant}.
$$
This type of flow is observed when the inner cylinder is driving (at low speeds).

When the outer cylinder is driving the motion one observes a flow that is very near that of a solid body. The usual analysis postulates the general flow, in the laminar regime,
$$
\vec v = ({a\over r^2}+b)(-y,x,0),
$$
with a pair of constant parameters that allows to satisfy the boundary conditions. In our notation,
$$
-\DD \Phi = {a\over r^2}(-y,x,0),~~~\vec X= bt(-y,x,0).\eqno(5.9)
$$
 
This flow satisfies the Euler-Lagrange equations (5.4) and (5.5) so long as $a$ remains constant but they do not restrict $b$ to be constant. The system
is thus underdetermined. But in the presence of viscosity Eq.(5.5) is replaced by (5.7)
and that gives the additional condition for a stationary flow
$$
\hat\mu\Delta v = 0 ~~~{\rm or}~~~b= {\rm constant}.  
$$
An effect of viscosity is to convert an underdetermined system to one that is well determined. This may seem fortuitous and unsatisfactory, but it is exactly the same
as what one finds by applying the Navier-Stokes equation. That approach also embodies the last equation and that too is undertermined when $\mu=0$.

Satisfying or not, this phenomenon is physically  significant, for it provides the explanation for why the flow, when not gradient type, tends to acquire a solid-body component,
a phenomenon that is observed in rotating superfluids as well, or at least in attempts
to analyze the same, as in the work of Feynman (1954a, 1954b) and of Landau and Lifshitz (1955). 

It remains to analyze the last of the Euler-Lagrange equations; suffice to say that it
is in close if not perfect agreement with the Navier-Stokes equation, for the Lagrangian (5.1) was chosen to make it so (Fronsdal 2014b). (The relative signs of the two quadratic terms in (5.6) is mirrored by the Navier - Stokes equation. It was this fact,
at first sight mysterious, that led to the proposed form of the Lagrangian.)

This Lagrangian density (5.1) is not  meant to be the last word; but to demonstrate that variational methods are available beyond the limitation to 
potential flows. The full ramifications of this will take some time to determine.

The decomposition of laminar, rotational flow velocity into two parts of which one is potential and the other is solid-body flow is precisely the same as in the traditional treatment with the Navier-Stokes equation. The equations of motion give a good account of the laminar flow observed at low rotation speeds. The more interesting (but more difficult) problem of the many types of flow that are observed at high rotation speeds have not yet been analyzed from this point of view. The application to superfluids,
with their two densities and/or two flows should be very interesting.

An  independent argument that leads to bivelocity flows in rarified gases has been advanced by Brenner (2013) and by Brenner et al (2013). But here the interpretation is very different.

There is an approach to turbulence, intimated by Onsager (see Eyink and Sreenavasen  2006)
and pursued in a very interesting paper by Lund and Regge (1976). They consider
a flow pattern  where the fluid velocity is irrotational everywhere except on a vortex line. They introduce
a function $\vec X$ from $\R^2$ to $\R^3$ that takes values at the coordinates of the vortex line. The time derivative $\dot{\vec X}$ is a vector field that is concentrated on the vortex line. It is thus intimately related to the vorticity. Subsequently, this field is generalized to extend to the whole space; it is natural to associate it with general vorticity. 

We suggest the following intuitive interpretation of our bi-velocity theory.  The
field $\vec v = \kappa \dot{\vec X}-\DD\Phi$ is the mass flow, associated with the transport of mass; the other component represents  ``vorticity flow". In the absence of viscosity the
vorticity flow is conserved; the primary effect of viscosity  is on vorticity.
\bb

\noindent{\bf VI. Relativistic hydrodynamics with unrestricted matter flows.}
\b

To include rotating bodies in General Relativity, we must upgrade the vector field $\vec X$ to a relativistic field.

A dramatic effect of relativization is that the fields become propagating.   In particular, propagation of the fields into empty space may be unavoidable, therefore it is prudent to
keep the number of propagating modes at a minimum. And that leads us to the antisymmetric tensor gauge field and its well known properties. This field has only spinless, propagating mode.

The 3-vector field  $\dot{\vec X}$ is the space part of the 4-vector with Cartesian components
$$
\tilde{dY^\mu} = \epsilon^{\mu\nu\lambda\sigma}Y_{\nu\lambda,\sigma}= (\dot {\vec X}, c\DD\cdot \vec X).
$$

\b

\ce{\bf  Special relativity. The antisymmetric tensor field.}

Let $Y = (Y_{\mu\nu})$ be an antisymmetric tensor field (a 2-form) and consider the Lagrangian density
$$
dY^2 = {c^2\over 4}g^{\mu\mu'}g^{\nu\nu'}g^{\lambda\lambda'}Y_{\mu\nu,\lambda}\sum_{\rm cyclic}Y_{\mu'\nu',\lambda'}.\eqno(6.1)
$$
Greek indices run over 1,2,3,0, latin indices over 1,2,3. The (inverse) metric tensor is the Lorentzian, diagonal with $g^{11}=g^{22} = g^{33} = -1, g^{00}=1/c^2$. It is invariant under the gauge transformation $\delta Y = d\xi$. In a 3-dimensional notation,
$$
X^i = {1\over 2}\epsilon^{ijk}Y_{jk}, ~~~\eta_i =  Y_{0i},
$$
The equation $\eta_i = \p_0\xi_i - \p_i \xi_0$  
can always be solved for the vector field $\vec \xi$;   there is a family of gauges in which the field $\vec \eta$ vanishes. 

In flat space, in terms of $\vec X$ and $\vec \eta$, 
$$
dY^2 = {1\over 2}\bigg(\dot{\vec X} + \DD\w \vec \eta\bigg)^2 - 
{c^2\over 2}(\DD\cdot \vec X)^2.
$$
The free field equations associated with this expression for the Lagrangian density are
$$
{d\over dt}\bigg( \dot{\vec X} + \DD\w\vec \eta\bigg) - c^2\DD\ (\DD\cdot \vec X) = 0,~~~\DD\w(\dot{\vec X}+\DD\w\vec \eta) = 0.
$$
The only propagating mode is a scalar mode. But we can add sources,  
$$
{d\over dt}\bigg( \dot{\vec X} + \DD\w\vec \eta\bigg) - c^2\DD(\DD\cdot \vec X) = \vec K,~~~\DD\w(\dot{\vec X}+\DD\w\vec \eta) = \vec K',\eqno(6.2)
$$
so that, in their presence, the field $\dot{\vec X}$ need not be  be irrotational.

  A stationary solution for the geometry of cylindrical Couette flow is $\vec X = bt(-y,x,0)$, ~$\DD\w \dot{\vec X} = 2b(0,0,1)$ and $\DD\cdot \vec X=0$; it is a solution of these field equations in a gauge where $\vec \eta=0$,
with $\vec K = 0$, but with 
$\vec K' = 2b(0,0,1).$ It is known (Barnett 1915) that a magnetic field is produced by a rapidly rotating cylinder made of iron and it is believed that the effect is general with an
effective magnetization proportional to the magnetic moment.  In the hope of
accounting for this effect, without a microscopic model, we add a gauge invariant 4-form
$$
\gamma Y\wedge F = {\gamma\over 4} \tilde Y^{\mu\nu}F_{\mu\nu} =\gamma(\vec X\cdot \vec E +c^{-1}\vec \eta\cdot \vec B),\eqno(6.3)
$$
where $\tilde Y$ is the dual,
$$
  \tilde Y^{\lambda\rho} =  {1\over 2}\epsilon^{\mu\nu\lambda\rho}Y_{\mu\nu},
$$
$F$ is the electromagnetic field strength, $F_{ij} = (\DD\w A)_{ij} = c^{-1}B_{ij}, F_{0i} = E_i$ and $\gamma$ is a parameter with dimension $g/sec \,cm^3$. This ties in nicely with speculations that have been voiced by several 
workers, of an intimate connection between solid body motion and electromagnetism.
The electric field may be related to measurements by Tolman (1910) and Wilson and Wilson (see Cullwick 1959, Rosser 1971).

The antisymmetric tensor field (the notoph) was first
studied by  Ogievetskij and Palubarinov (1964). The term (6.3) represents a mixing that can be interpreted as giving a mass to the photon. A preliminary investigation suggests that this effective photon mass is too small for direct experimental detection. However, in the presence of large, rapidly rotating galaxies the total effect on the propagation of electromagnetic radiation may become significant, providing an explanation for all or 
a part of the observed anomalous gravitational lensing. We may fancy a direct connection between the $Y$-field and Dark Matter. The antisymmetric tensor field also appears in string theory, as the Kalb-Ramond field (Kalb and Ramond 1974), or the $B$-field, and in that context (6.3) is known as the Green-Schwarz term.

The velocity $\dot{\vec X}$ is responsible for turbulence. The case of vortex strings
where the vorticity is concentrated on a line  was studied by Lund and Regge, as already mentioned.   They   promoted it  to a global field of the
same type as the notoph field. But to define the theory so as to get the requisite number of degrees of freedom we must impose the constraint.  The association between the notoph field and turbulence is thus not new here.


The propagating, scalar mode is a new form of matter that interacts with gravity and with the electromagnetic field strength. Being endowed with electric and magnetic polarizability, it would 
make a contribution to Dark Matter.

The term $\kappa \rho  \dot{\vec X}\cdot\DD\Phi$ can be made invariant under Lorentz transformations by expanding it to
$$
\kappa \rho{c^2\over 2}\epsilon_{\mu\nu\lambda\rho}Y_{\mu\nu,\lambda}\psi_{,\rho}.\eqno(6.4)
$$
Here $\psi$ is a Lorentz scalar;  if $\rho$ is uniform it is a boundary term.

Introduce the dual as in (6.3),
Then
$$
{c^2\over 2}\epsilon_{\mu\nu\lambda\rho}Y_{\mu\nu,\lambda}\psi_{,\rho}
= c^2\tilde Y_{\lambda\rho,\lambda}\psi_{,\rho}    
=\tilde Y_{0i,0}\psi_{,i} + \tilde Y_{i0,i}\psi_{,0} - \tilde Y_{ij,i}\psi_{,j},
$$
$$
 ~~~\tilde Y_{i0,i} = \DD\cdot \vec X,~~~\tilde Y_{ij,i} = (\DD\w \vec \eta)_j,
$$
and
$$
{c^2\over 2}\epsilon^{\mu\nu\lambda\rho}Y_{\mu\nu,\lambda}\psi_{,\rho}
= \dot{\vec X}\cdot \DD\psi - (\DD\cdot\vec X)\dot\psi +(\DD\w \vec \eta)\cdot \DD\psi.\eqno(6.5)
$$
The first term on the right is the one that appears in the non relativistic theory,
the term that we are trying to promote to the relativistic context. The second term
vanishes in the case of the elementary solution (5.9). The last term 
vanishes in  physical gauge  ($\vec \eta = 0$). 

Now we consider the total Lagrangian density
$$
\L = {\rho\over 2}(g^{\mu\nu}\psi_{,\mu}\psi_{,\nu}-c^2) + \rho{c^2\over 4}dY^2  +\rho\kappa {c^2\over 2}\epsilon^{\mu\nu\lambda\rho}Y_{\mu\nu,\lambda}\psi_{,\rho} + \gamma YF
+{1\over 8}F^2   -f -sT\eqno(6.6)
$$
and the action $\A = \int d^3xdt\,\L$. 
The first term is the Lorentz invariant contribution that appears in the non relativistic approximation as $\rho(\dot\Phi - \DD\Phi^2/2).$
 
The variation of the action with respect to $\psi$ is $-\delta\psi$ times
$$
{d\over dt}\bigg(\rho({\dot\psi\over c^2}-\DD\cdot \vec X)\bigg)  +
\DD\cdot\bigg(\rho(\dot{\kappa \vec X} -\DD\Phi  + \DD\w \vec\eta)\bigg) =: {d\over dt}J^0 
+ \DD\cdot \vec J.\eqno(6.7)
$$
In the non relativistic limit $\dot\psi/c^2$ is unity, and for the stationary rotations considered $\vec X$ is divergence-less.  The boundary conditions require that the current $\vec J$  be normal
to the boundary.

The variation of the action with respect to the field $\vec X$ is
$$
\int d^3x dt \,\delta\vec X\cdot \bigg({d\over dt}\big(\rho( \dot{\vec X} + \kappa\DD\Phi+\DD\w\vec \eta)\big) + 
\DD\big(\rho(\dot\psi-c^2\DD\cdot \vec X)\big) -\gamma \vec E\bigg).\eqno(6.8)
$$
Setting this to zero gives the field equation
$$
{d\over dt}\big(\rho( \dot{\vec X} + \kappa \DD\Phi+\DD\w\vec \eta)\big) + 
\DD\big(\rho(\dot\psi-c^2\DD\cdot \vec X)\big) =\gamma \vec E.\eqno(6.9)
$$
If we want the stationary, solid-body rotating motion of the non relativistc theory
to be a solution of the relativistic theory, then we may need to an electric field. 
 An electric field is expected on the basis of an experiment by Tolman (1910). It may also have something to do with the anomalous Seebeck effect.
But at this time we are far from understanding all the ramifications of this 
interaction with the electromagnetic field. That an unexpected role may be played
by the electric field in connection with vortex motion, or that an electromagnetic analogy may be seen here, was suggested by Feynman in connection with liquid helium (Feynman 1954b page 24).

  An unexpected effect of moving up to the relativistic context is an additional field equation related to the variation of the action with respect to the gauge field $\eta$. The gauge is fixed in the non relativistic theory but variation of the gauge field nevertheless furnishes an additional Euler-Lagrange equation.
The variation of the action with respect to the field $\vec \eta$ is
$$
\int d^3x dt \,\delta\vec \eta\cdot \bigg(
\DD \w\big(\rho(\dot{\vec X} +\kappa \DD\Phi+\DD\w\vec \eta\big) -{\gamma\over c}\vec B
\bigg).
\eqno(6.10)
$$
Setting this to zero gives the field equation (in the physical gauge, $\eta = 0$)
$$
\DD \w\big(\rho(\dot{\vec X} +\kappa\DD\Phi\big) =
{\gamma\over c}\vec B.
\eqno(6.11)
$$
This is  a constraint on a gauge theory that is difficult to understand within the gauge-fixed, non-relativistic context.

In the special case of an incompressible fluid, with $\rho$ uniform, and $\vec B = 0$, this constraint would make the curl of $\dot{\vec X}$ equal to zero, which would throw us back on potential theory. But a closer examination gives a different result. For example,
suppose that the flow is of the symmetric type (5.9), with the magnetic field   $2bc(0,0,1)$, then (6.11) is a constraint on the density, solved by
$$
\rho = {\gamma b r^2 + C\over b r^2 + \kappa a},~~~~ C~{\rm constant}.
$$

The appearance of the magnetic field may be related to the Barnett effect.
Barnett, in an extensive series of experiments, discovered that a rotating iron cylinder developed a magnetic field, unrelated to any microscopic magnetic moment  (Barnett 1915). A direct association between magnetism and turbulence was predicted by Batchelor (1950). This interaction has been 
used to explain the reduction of angular momentum in galactic disks (Balbus and Hawley 1991)). 

Work in progress includes solving Maxwell's equation but that will not be reported here.

At this time we are far from understanding all the ramifications of this 
interaction with the electromagnetc field. 
The action of the Galilei group in   non-relativistic hydrodynamics (see ``Remark" at the end of 
section III.3) comes from the action of the Lorentz group on the scalar field  $\psi$. The Lorentz group acts on the antisymmetric field $Y$ as well,
but in the physical gauge ($\vec \eta = 0$) it does not generate an affine variation of
the vector field $\dot{\vec X}$. To put it simply, if inaccurately, the Galilei group does not act on this vector field. 
\bb
\ve

\ce{\bf  General Relativity. The energy momentum tensor}

The Lagrangian density (6.6) is easily generalized to the case of an arbitrary space time metric.  The matter action is
$$
{\cal A} =  \int d^4x\sqrt{-g} \L,
$$
with
$$
\L = {\rho\over 2}(g^{\mu\nu}\psi_{,\mu}\psi_{,\nu}-c^2)  +{1\over 8}g^{\mu\mu'}g^{\nu\nu'}F_{\mu\nu}F_{\mu'\nu'} +\gamma YF
$$
$$
+ \rho{c^2\over 4}g^{\mu\mu'}g^{\nu\nu'}g^{\lambda\lambda'}Y_{\mu\nu,\lambda}
\sum_{\rm cyclic}Y_{\mu'\nu',\lambda'}  +\rho\kappa{c^2\over 2}\epsilon^{\mu\nu\lambda\rho}Y_{\mu\nu,\lambda}\psi_{,\rho} 
  -f -sT. 
$$
The associated energy momentum tensor is
$$
T_{\mu\nu} = 2{\delta {\cal \L}\over \delta g^{\mu\nu}}-\L g_{\mu\nu} 
$$
and Einstein's equations take the form 
$G_{\mu\nu}= 8\pi G T_{\mu\nu}$.

The Bianchi constraint on $T$ is now satisfied by virtue of the field equations.
Another report will include solutions of the field equation in a fixed metric field;
the special case of the Kerr metric is especially interesting.

\b

\ce{\bf Final remarks} 

It is interesting to look back, from the current perspective, at the Tolman formula for the energy momentum tensor. If we admit that the familiar, 3-dimensional velocity field is to be promoted to a timelike 4-vector field, then we find no fault with the formula (2.1), as an example; it has ample physical application. (For the normalization condition  
(2.2) there can be no justification,  since it implies the loss of the equation of continuity.)  But to argue, as Tolman does (Tolman 1934), and as Minkowski did more than 100 years ago, (Minkowski 1908), that the passage from non relativistic hydrodynamics to a relativistic theory must include replacing the familiar 3-vector velocity field with a time-like 4-vector field, is unwarranted. It is contradicted by the system with Lagrangian density (6.6), where the non relativistic velocity is the time derivative of a vector field (and thus a 3-vector in the sense of 3-dimensional differential geometry) and the relativistic version is the dual of an exact 3-form.
(It is sometimes argued that the  answer is to promote the potential $\vec X$ to a four vector and to define `velocity' as the derivative with respect to proper time.
But the concept of proper time has no place in a theory of fields over space time.)

Several other applications come to mind. Superfluids, with multiple densities and/or
multiple velocity fields is perhaps the most interesting.  A review of the theory of instabilities of Couette flow will be a challenging test of the theory and is therefore  urgent. A fresh approach to the theory of electrodynamics of fluids could begin with an application to electrolysis. An application to galaxies rotating in Kerr space time is in progress. 
The impact on rotations in plasma physics is currently a vital subject and yet another field of application.

\bb

{\bf Acknowledgements.} I am grateful to Tom Wilcox and 
Evgenyi Ivanov   for stimulating conversations. I thank NTNU in Trondheim and the Bogoliubov Institute in Dubna for
hospitality. Special thanks are due Volodya Kadyshevsky for a  wonderful period
spent at JINR. His passing away on September 24, 2014 is a great personal loss.
I thank a referee for useful remarks.
\bb
 \bb 
 
{\bf References}

\noindent Andreassen, H., ``The Einstein-Vlasov System/Kinetic Theory", 

	Living Rev. Rel. {\bf 14} 4-55  (2011). 
	
\noindent	Balbus, S.A. and Hawley, J.F., ``A powerful local shear instability in weakly 

magnetized disks. I - Linear analysis", ApJ {\bf 376}, 214-233 (1991).
  
\noindent Bardeen, J.M., ``A variational principle for rotating stars
in General Relativity", 

 Astrophys. J. {\bf 162}, 71-95 (1970).
 
 \noindent Barnett, S.J., ``Magnetization by rotation", Phys.Rerv, {\bf 6} 239-270 (1915).
 
 \noindent Batchelor,G.K., ~~``On the spontaneous magnetic field in a conducting liquid in 
 
 turbulent motion", Proc.R.Soc., {\bf 201} 405-416 (1950).
 
 \noindent Bogoliubov, N.N., (1947). "On the Theory of Superfluidity",
 Izv. Academii 
 
 Nauk USSR (in Russian) {\bf 11} (1): 77, 1947. English translation in Journal of 
 
 Physics {11} (1): 23?32, 1947.
 
\noindent Brenner, H. ``Proposal of a critical test of the Navier-Stokes-Fourier paradigm 

for compressible fluid continua",  Phys. Rev. {\bf E 87}, 013014 (2013).
 
 \noindent Brenner, H., Dongari, N., White, C., Ritos, K.,  Dodzie, S.K, and Reese, J.M.,

`` A molecular dynamics test of the Navier-Stokes-Fourier equations in 

compressible gaseous continua", MIT preprint 2013.
  
\noindent Chandrasekhar, S.,  ``The Highly Collapsed Configurations of a Stellar Mass" 

(second paper), MNRAS {\bf 95}, 207-25 (1935).

\noindent Cullwick, E.G., {\it Electromagnetism and Relativity}, Longmans, London, 1959, 

pp. 161-174.

\noindent Courtright, T. and Fairlie, D. ``A Galileon Primer", arXiv 1212 6972v1 (1212).

J. Cosmol. Astropart. Phys.JCAP05(2010)015(arXiv:1003.5917[hep-th]. 

\noindent de la Mora, J.F. and Puri,  A., ``Two fluid Euler theory of sound dispersion in  

gas mixtures of disparate masses", J. Fluid Mec. {\bf 168} 369-382 (1985).

\noindent de Rham, C. and Tolley, A.J., ``2010 DBI and Galileon reunited,"

J. Cosmol. Astropart. Phys. JCAP05(2010)015 (arXiv:1003.5917[hep-th])

\noindent  Eddington, A.S., {\it The internal constitution of stars}.   

Cambridge University Press 1926. 

\noindent Ellis, G. F. R., ``Relativistic cosmology,
General relativity and cosmology", 

Proc. Internat. School of Physics Enrico Fermi",
Italian Phys. Soc., 

Varenna, 1969, pp. 104{182. Academic Press, New York, 1971.

\noindent Eyink,  G.L. and Sreenivasan, ``K.R. Onsager and the theory of hydrodynamic 

turbulence", Rev. Mod. Phys., Volume 78, January (2006).

 \noindent Fairlie, D., ``Comments on Galileons", 
 
 Phys. A: Math. Theor. {\bf 44}
(2011) 305201 (9pp)doi:10.1088/1751-8113/44/30/305201

\noindent Fetter, A.L. and Walecka, J.D., {\it Theoretical Mechanics of Particles

and Continua},  
MacGraw-Hill NY 1980.

\noindent Feynman, R.P., `` Atomic theory of two-fluid model of liquid helium", 

Phys.Rev. {"\bf 94} 262-277 (1954a).

\noindent Feynman, R.P., {\it Progress in Low Temperature Physics}, Vol. 1, 17-53,

North Holland, Amsterdam, 17-53 (1954b).

\noindent Fronsdal, C., ``Ideal Stars and General Relativity", 

Gen.Rel.Grav.  {\bf 39} 1971-2000 (2007a). 

\noindent Fronsdal, C., ``Reissner-Nordstr\"om and charged polytropes",

Lett.Math.Phys. 82, 255-273 (2007b).

 \noindent Fronsdal, C., ``Stability of polytropes", Phys.Rev.D, 104019 (2008).




 
\noindent Fronsdal C. and Wilcox, T.J. ``An equation of state for the dark matter" , 

arXiv 1106.2262  (2011).



\noindent Fronsdal, C. , ``Heat and Gravitation. The action Principle", 

Entropy 16(3),1515-1546 (2014a).

\noindent Fronsdal, C., ``Action Principle for Hydrodynamics and Thermodynamics 

including general, rotational flows"}, 

arxiv 1405.7138v3[physics.gen-ph](v3 dated 19 June 2015).

\noindent Fronsdal, C. , {\it Thermodynamics of fluids}, in progress (2014c), 

fronsdal.physics.ucla.edu.   

\noindent  Gibbs, J.W., ``On the equilibrium of heterogeneous substances,"

Trans.Conn.Acad. Vol 3, 108-524 ( 1878). 

\noindent Israel W.  and J. M. Stewart, J.M., ``Transient Relativistic Thermodynamics and 

Kinetic Theory", Annals Phys. 118, 341- 372 (1979). 

\noindent Kalb, M. and Ramond, P. (1974). ``Classical direct interstring action". 

Phys. Rev. {\bf D 9} (8): 2273-2284 (1974).

\noindent Landau, L. D., and Lifshitz, E. M., {\it Statistical Physics, Pergamon Press}, (1955), 

\noindent Landau, L. D., and Lifshitz, E. M., {\it Statistical Physics, Pergamon Press}, (1958), 

Chapter 5.

\noindent Lund,  F. and Reggee T.,  ``Unified Approach to strings and vortices with soliton 
 
solutions", Phys. Rev. D. {\bf 14} 1524-1548 (1976).

\noindent Mannheim, P.D., "Alternatives to dark matter and dark energy",

Progress in Particle and Nuclear Physics, {\bf 56} 340-445  (2006).

\noindent Mannheim, P.D., O'Brien, J. G. and
Cox, D.E.,  
``Limitations of the standard 

gravitational perfect 
fluid paradigm", 

Gen. Relativity Gravitation {\bf 42}   2561-2584 (2010).

\noindent Marsden, J.E. and Weinstein, A. ``The hamiltonian structure of the Maxwell.-

Vlasov equations", Physica {4D} 394-406 (1982).

\noindent  Massieu, F.,  ``Thermodynamique: M\'emoire sur les fonctions caract\'eristiques 

des divers fluides et sur 
 la  th\'eorie des vapeurs",   
  Comptes Rendues de 
  
  l' AcadŽ. des Sciences 1876.  M\'emoires pr\'esent\'es \`a 
  l'Acad\'emie des Sciences 
  
  par des Savants Etrangers, XXII, 1-92 (1876).
 

\noindent Misner, W., Thorne K.S. and Wheeler, J.A., ``Gravitation", W.H. Freeman, 

N.Y. 1972.



\noindent Morrison, P.J. "Bracket formulation for irreversible classical fields", 

Physics Letters {100A}, 423-427 (1984).

\noindent Ogievetskij, V.I. and Palubarinov, ``Minimal interactions between spin 0 and 

spin 1 fields",
J. Exptl. Theoret. Phys. (U.S.S.R.) {\bf 46} 1048-1055 (1964).

\noindent Okabe, T.,  Morrison, P.J., Friedrichsen, J.E. III and Shepley L.C., ``Hamiltonian 

dynamics of spatially-homogeneous Vlasov-Einstein systems",  

Phys.Rev. {\bf D 84} 024011 (2011).



\noindent  Oppenheimer, J.R. and Volkoff, G.M., ``On Massive Neutron Cores", 

Phys. Rev. {\bf 55}, 374-,381 (1939).

\noindent Pavlov, V.P. and Sergeev, V.M., ``Thermodynamics  from the differential 

geometry standpoint", Theor. Math. Phys., {\bf 157}, 141-148 (2008).

\noindent Pellegrini, G.N. and Swift, A.R.,
``Maxwell's equations in a rotating medium: 

Is there a problem?",
 Am. J.  Phys.{\bf 63} 694 ,1995; doi: 10.1119/1.17839.
 
 \noindent Poincar\'e, H. {\it Thermodynamique}, Gauthier-Villars, Paris 1908.

\noindent Rosser, W.G.V., {\it An introduction to the theory of Relativity}, Butterworths,

 London, 1971, pp. 341-345.
 
\noindent  Rowlinson,  {\it Liquids and liquid mixtures}, Butterworths, London 1959.
 
\noindent  Saha,  M.N., ``On a physical theory of stellar spectra", 
  
  Proc.Roy.Soc.Series A, {\bf 99}, 135-153 (1921).
 
 \noindent Schaefer and Teaney, Rept. Prog. Phys. 72 (2009) 126001.
 
\noindent Schutz, B.F. Jr., ``Perfect fluids in General Relativity, Velocity potentials and a 

variational principle", Phys.Rev. {\bf D2}, 2762-2771 (1970).

\noindent  Stoner, E.C. (1930), Philosophical magazine, {\bf 9} (60), 944-963 (1930).

\noindent  Taub, A.H., ``General Relativistic Variational Principle for  Perfect Fluids", 

Phys.Rev. {\bf 94}, 1468-1470 (1954).

\noindent Tolman, R.C., {\it Relativity, Thermodynamics and Cosmology},

Clarendon, Oxford 1934.

\noindent Tolman, R.C., ``The electromotive force produced in solutions by

centrifugal action", 
Phys.Chem. MIT, {\bf 59}, 121-147 (1910).

\noindent Verlinde, E. P., ``On the Origin of Gravity and the Laws of Newton",

JHEP 1104:029,2011, arXiv:1001.0785. 


\noindent Weinberg, S., ``Gravitation and Cosmology: Principles and Applications of the 

General Theory of Relativity", John Wiley, N.Y. 1972.

\noindent Wilson, H.A., ``On the Electric Effect of rotating a Dielectric in a Magnetic 

Field,"
Phil.Trans.,  {\bf  A 204}  1904.
 
\noindent Wilson , M. and Wilson, H.A., ``On the electric effect of rotating a magnetic 

insulator in a magnetic field ", Proc.R.Soc. London Ser.A {\bf 89},99-106, 1913.

\ve

\end{document}